\documentclass[pre,showpacs,preprint,superscriptaddress]{revtex4}
\usepackage{amssymb,graphicx,amsbsy,amsmath}

\begin{document}
\title{Ising model on a hyperbolic plane with a boundary}
\author{Seung Ki Baek}
\affiliation{Integrated Science Laboratory, Ume{\aa} University, SE-901 87
Ume{\aa}, Sweden}
\author{Harri M\"akel\"a}
\author{Petter Minnhagen}
\affiliation{Department of Physics, Ume{\aa} University, SE-901 87 Ume{\aa},
Sweden}
\author{Beom Jun Kim}
\email[Corresponding author, E-mail:]{beomjun@skku.edu}
\affiliation{BK21 Physics Research Division and Department of Physics,
Sungkyunkwan University, Suwon 440-746, Korea}

\begin{abstract}
A hyperbolic plane can be modeled by a structure called the enhanced binary
tree. We study the ferromagnetic Ising model on top of the enhanced binary
tree using the renormalization-group analysis in combination with
transfer-matrix calculations. We find a reasonable agreement with Monte
Carlo calculations on the transition point, and the resulting critical
exponents suggest the mean-field surface critical behavior.
\end{abstract}


\maketitle

Critical phenomena on a plane has been one of the most well-studied area
in statistical physics. One obvious reason is that it is a plane that we
can most easily visualize, but it is also because a two-dimensional (2D)
system is often analytically tractable. It is true that one could say the
same about one-dimensional (1D) systems as well, but collective orders in a
low-dimensional system tend to be fragile against thermal disorders; it is
from the dimensionality $d=2$ for many model systems to exhibit a
phase transition at a nonzero temperature $T>0$. These remarks are best
illustrated by the Ising model defined by the following Hamiltonian
\begin{equation}
H = -J \sum_{\left< ij \right>} S_i S_j - h \sum_i S_i,
\label{eq:ham}
\end{equation}
where $J>0$ is the ferromagnetic interaction strength, and $h$ is the
magnetic-field strength. The first summation is over
every pair of nearest neighbors, and spin at site $k$ can take its value $S_k$ from
$\pm 1$. As is well known, the 1D Ising model does not have any magnetic order
at finite temperatures, while the 2D counterpart undergoes a continuous
order-disorder transition at a finite coupling strength $K \equiv \beta J$.
Here the inverse temperature is denoted as $\beta \equiv (k_B T)^{-1}$,
where $k_B$ is the Boltzmann constant~\cite{pl}. In this regard, the lower
critical dimension of the Ising model is $2$.

By planes, however, we do not always have to mean the flat geometry.
There are many kinds of curved planes observable in physical or biological
structures, and critical phenomena on such planes may exhibit intriguing
features. One can assign the Gaussian curvature to a plane, which is
positive (negative) when the plane looks like a part of a sphere (saddle).
With a constant positive Gaussian curvature, the plane will be eventually
closed to form a sphere, whose radius is inversely proportional to the size
of the curvature. This implies that the curvature should vanish at every
local point if we are to work with a very large system size. Even though
there can remain global topological constraints determined by the positive
curvature, many statistical-physical properties will converge to those of
the flat geometry in the large-size limit. In the hyperbolic geometry with
a negative constant curvature, on the other hand, the magnitude of the
curvature is not necessarily coupled to the system size, and therefore this
case is often regarded as more suitable to study effects of the curvature.
The price is that the surface area $\mathcal{A}$ expands exponentially as
its radial length scale grows. It immediately leads the boundary of the
surface $\partial \mathcal{A}$ to expand at the same rate, so we find that
the boundary fraction $\partial \mathcal{A} / \mathcal{A}$ never vanishes
even in the large-size limit. The thermodynamic limit is not uniquely
defined for this reason. For example, one may use the periodic boundary
condition~\cite{sausset} and the behavior will not necessarily be the same
as with the open boundary condition.
By neglecting the presence of the boundary, it has been argued
by many authors that the Ising model will undergo a mean-field-like phase
transition on a negatively curved plane: a Monte Carlo analysis away from
the boundary suggested convergence to the mean-field exponents in
Ref.~\cite{shima}, which was later supported by the corner transfer-matrix
renormalization-group (RG) method~\cite{ueda}. The Ginzburg-Landau theory
provides a qualitative explanation for this mean-field behavior when only
the bulk part is considered~\cite{qclock}. For rigorous results under
transitivity, one may refer to Ref.~\cite{lyons} and references
therein.

\begin{figure}
\includegraphics[width=0.25\textwidth]{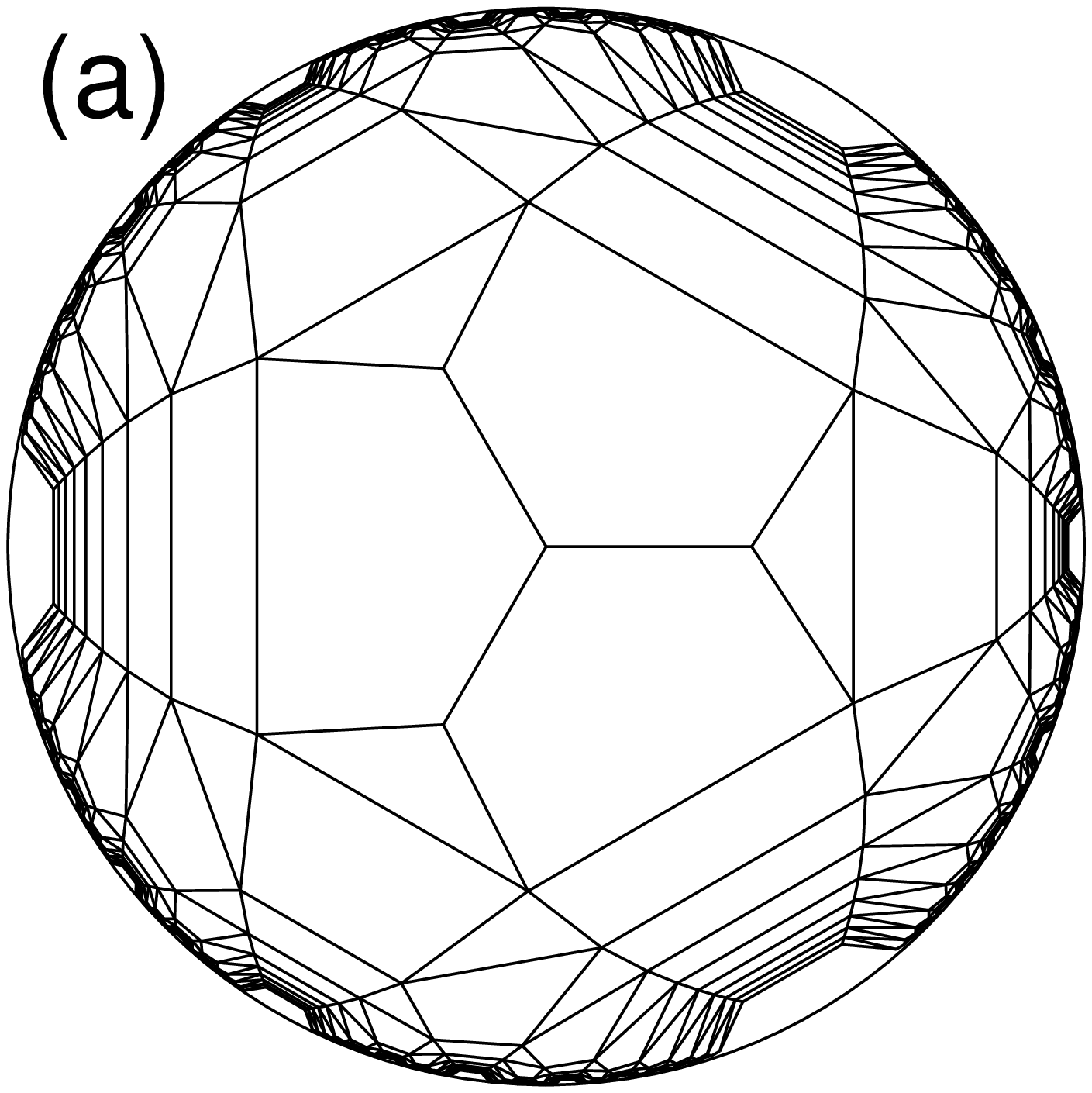}
\includegraphics[width=0.25\textwidth]{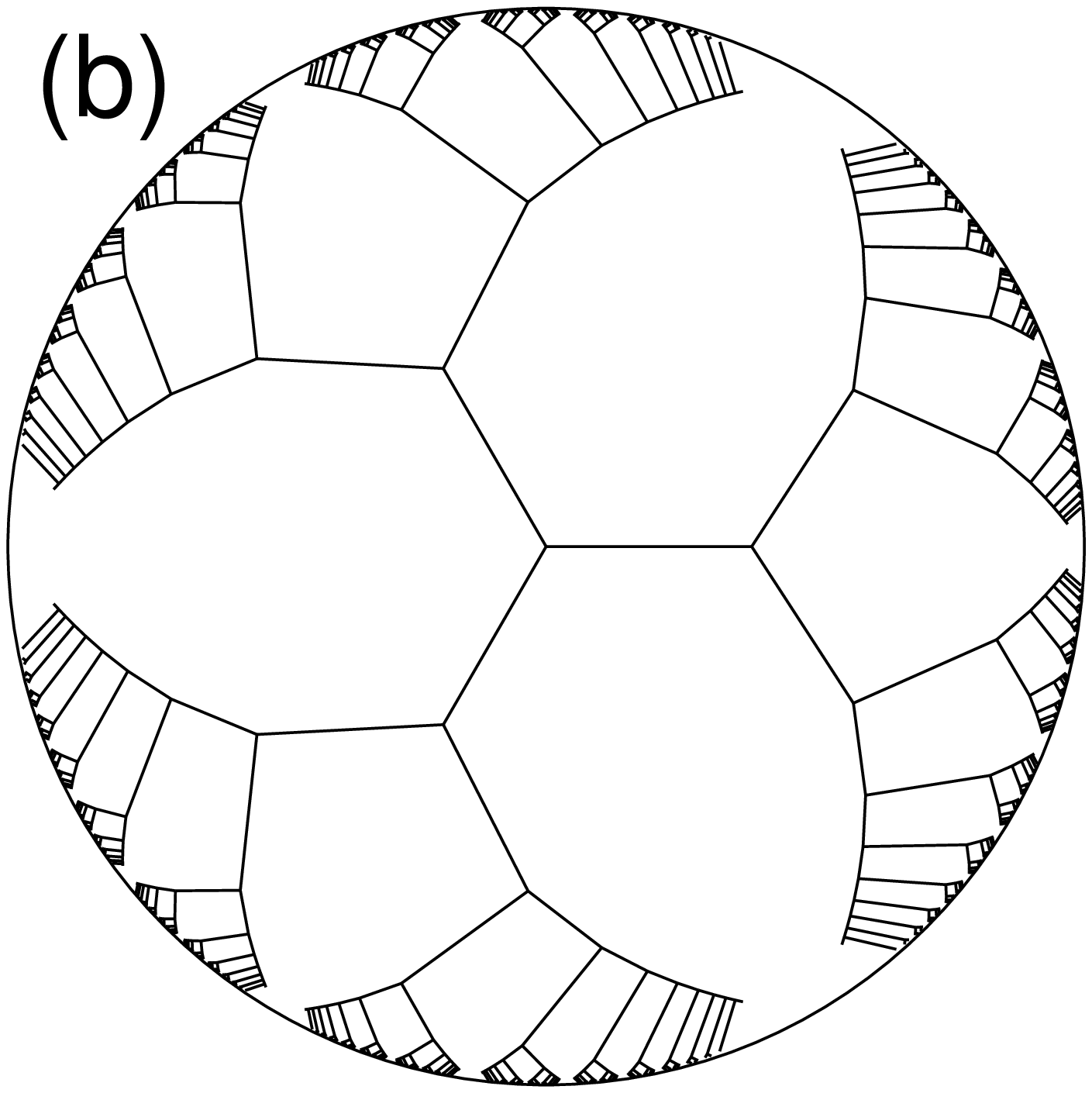}
\includegraphics[width=0.43\textwidth]{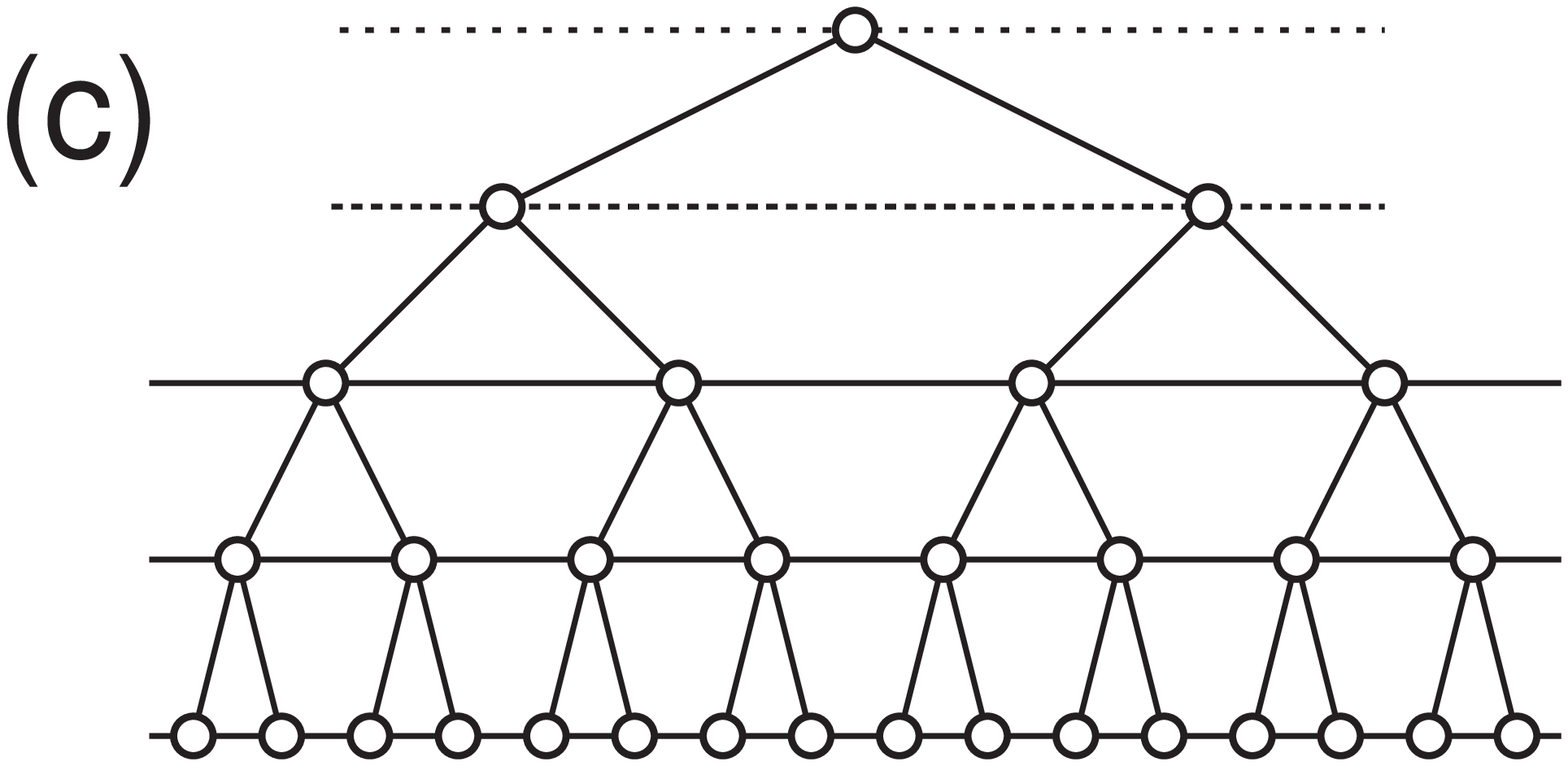}
\caption{Schematic descriptions of (a) the EBT and (b) the simple binary
tree, drawn on the Poincar{\'e} disks. The number of layers is $L=10$ in
both the cases. (c) The actual structure used in our MC calculations. Here
$L=4$ is shown as an example, and the structure is extended further down for
larger $L$.
In (c), we remove the two upper most horizontal connections since they
either make a self-link (the top dotted line) or double links (the dashed
lines) under the periodic boundary condition in the horizontal direction.
}
\label{fig:ebt}
\end{figure}

By accepting this nonvanishing boundary as a part of physics, it
becomes possible to study nontrivial statistical-physical properties due to
the boundary, as has been done in Refs.~\cite{melin,auriac,perc,qclock}. A
commonly used approach to simulate such a plane is to begin with hyperbolic
tessellation with regular polygons and truncate the lattice generation at a
certain layer~\cite{shima,shima2006a}. An advantage is that it makes every
point equivalent except at the boundary layer. But there is also an
alternative lattice structure for such a plane, called the enhanced binary
tree (EBT) as shown in Fig.~\ref{fig:ebt}(a)~\cite{nogawa}. It is obtained
by adding links between branches in a binary tree [Fig.~\ref{fig:ebt}(b)],
so the EBT itself is not a tree, strictly speaking. Although it is not a
uniform tiling of a hyperbolic plane but made up of triangles and
tetragons, this structure is actually easier to
study analytically: for the bond-percolation problem, for example, it is
possible to argue that the emergence of a single giant cluster occurs at a
critical occupation probability $p = p_c = 1/2$~\cite{an,bnd}. This analytic
tractability is particularly important because numerical calculations suffer
from the exponential growth of the system size and therefore can only give
very rough estimates. There can be in fact one more transition point where
the correlation diverges in percolation or the Ising
model~\cite{perc,qclock}, but in this work, we focus on the order-disorder
transition where the order parameter becomes nonzero. The simple binary tree
in Fig.~\ref{fig:ebt}(b), for example, cannot have the latter type of
order-disorder transition at any finite temperatures~\cite{qclock}.

This Brief Report is intended to extend the analytic approaches for
percolation to the Ising model on the EBT. We mainly rely on real-space RG
methods and compare the results with the numerical data obtained by Monte
Carlo (MC) calculations. Combined with the transfer-matrix method, this
approximate RG calculation allows us to estimate the critical point and
critical indices as well. The results suggest the mean-field critical
behavior of free surfaces.

\begin{figure}
\includegraphics[width=0.45\textwidth]{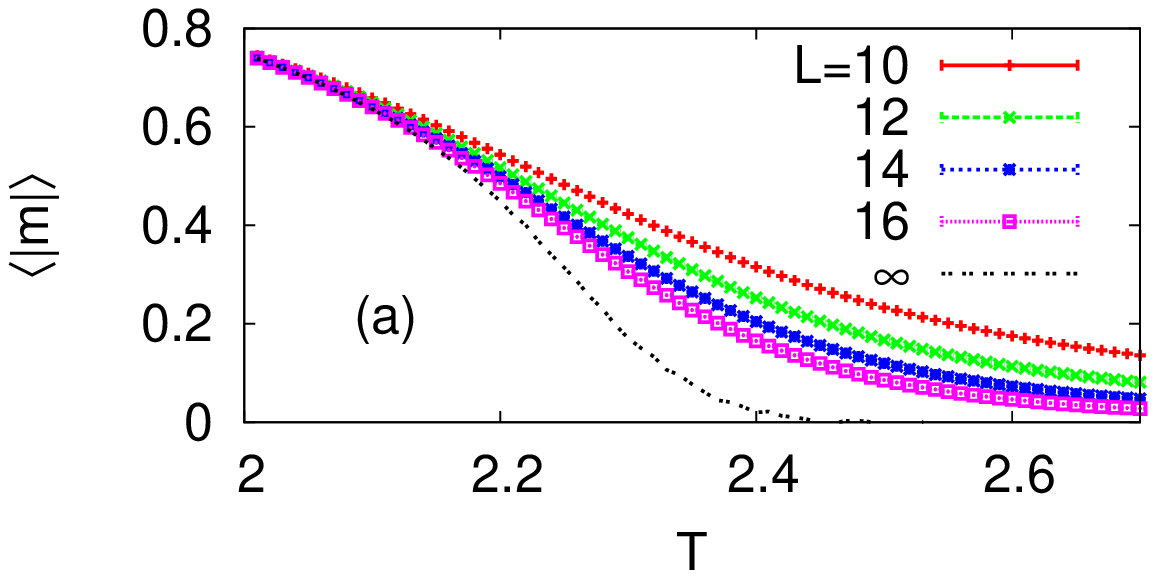}
\includegraphics[width=0.45\textwidth]{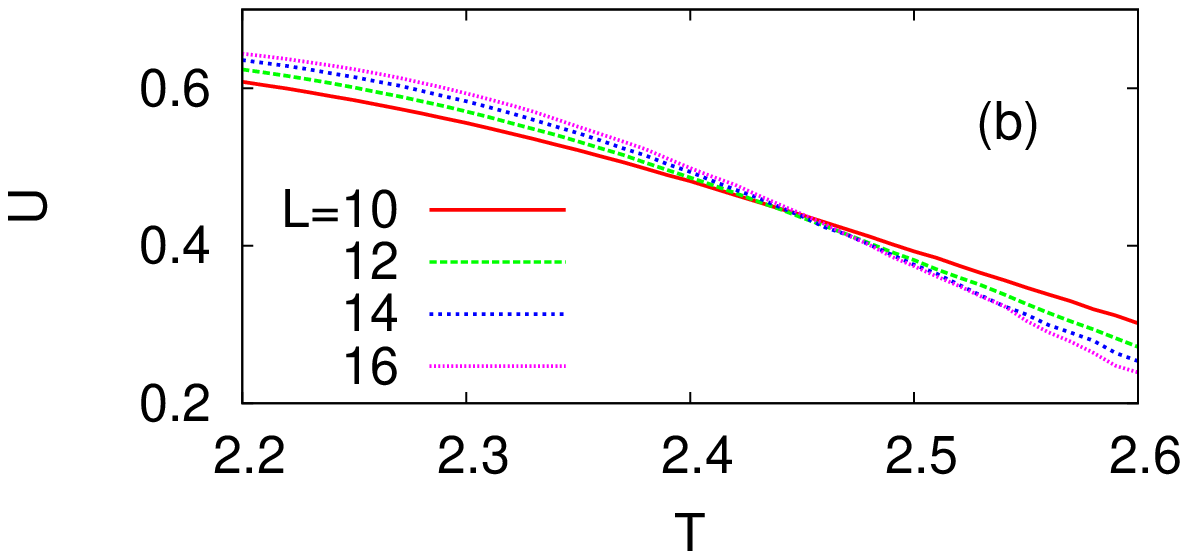}
\caption{(Color online) (a) Magnetic order parameter and (b) Binder's
cumulant of the Ising model on the EBT structures with different sizes.
The symbol $\infty$ in panel (a) means extrapolated values according to
Eq.~(\ref{eq:ext}).}
\label{fig:mc}
\end{figure}

Let us begin with numerical methods used in this work, and give a very rough
estimate of the critical coupling $K_c$ of the EBT. For MC calculations, an
EBT structure is constructed with a certain number of layers, $L$, as shown
in Fig.~\ref{fig:ebt}(c), where a periodic boundary condition is imposed in
the horizontal direction.
The system size is then given as $N = 2^{L+1}-1$,
which is an exponential function of $L$ as mentioned above. The Ising
Hamiltonian [Eq.~(\ref{eq:ham})] is simulated by the Wolff single-cluster
algorithm~\cite{wolff}. The magnetic order parameter is defined by
\[ \left<|m|\right> \equiv \left< \left| \frac{1}{N} \sum_i^N S_i \right|
\right>, \]
where the bracket $\left< \ldots \right>$ means the thermal average
[Fig.~\ref{fig:mc}(a)]. A convenient quantity to locate the critical point
is Binder's cumulant defined by
\[ U \equiv 1 - \frac{\left< | m |^4 \right>}{3\left< | m |^2 \right>^2}, \]
whose crossing point suggests $T_c \approx 2.45(2)$ in units of $J/k_B$
[Fig.~\ref{fig:mc}(b)].
We extrapolate the magnetic order parameter under the assumption
that~\cite{perc}
\begin{equation}
\left< |m| \right> \sim a N^{-\phi} + b
\label{eq:ext}
\end{equation}
where the parameters $a, b$, and $\phi$ are found by the least-square fitting
at each $T$. That is, we choose the fitting parameter $\phi$ that best
describes $\left< |m| \right>$ as a linear function of $N^{-\phi}$.
We then observe that the limiting value of $\left< |m| \right>$
at $N \rightarrow \infty$ could vanish at $T \gtrsim 2.4$
[Fig.~\ref{fig:mc}(a)], which supports the above estimation of $T_c$. The
exponent $\phi$ is found to have $\phi \approx 0.2$ around $T_c$.

\begin{figure}
\includegraphics[width=0.33\textwidth]{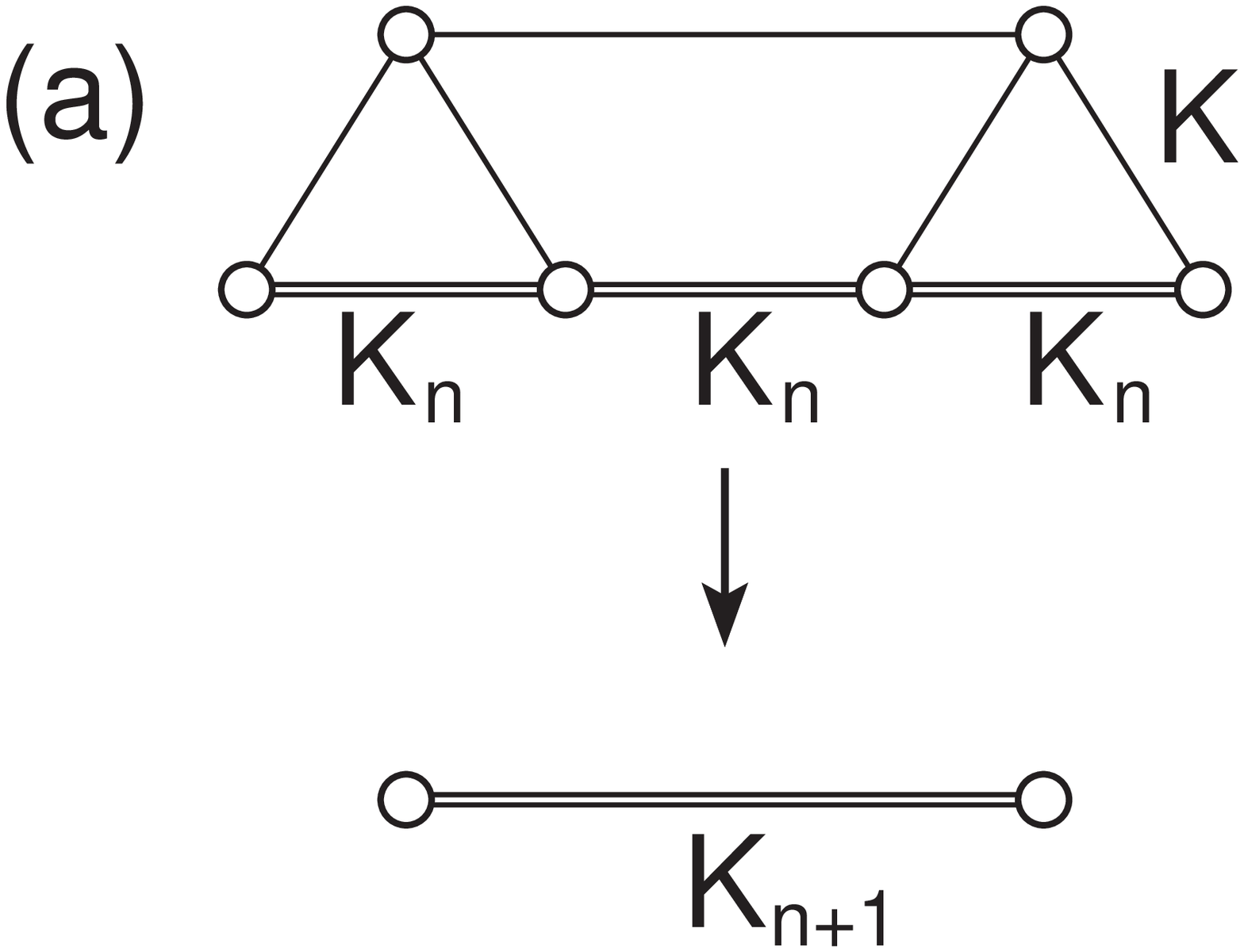}
\includegraphics[width=0.45\textwidth]{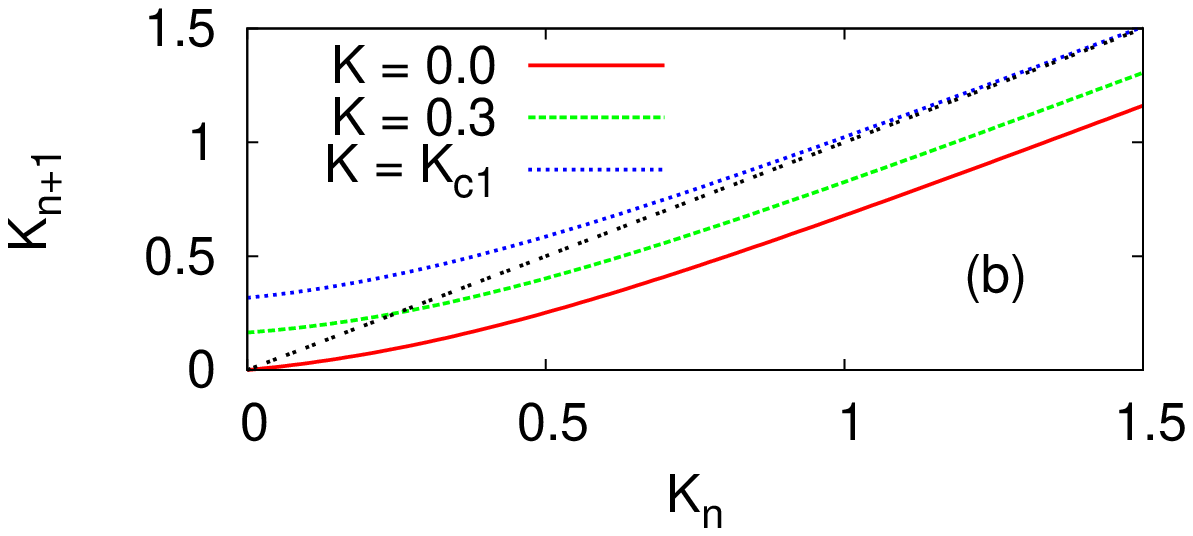}
\caption{(Color online) (a) Recursion scheme based on the finite-lattice
method. The double lines represent coarse-grained effective couplings.
(b) Iteration by Eq.~(\ref{eq:recur}) at various $K$'s.
The straight line has slope $1$, i.e., $K_{n+1} = K_n$.
}
\label{fig:spn}
\end{figure}

In order to study the problem analytically, let us consider the block-spin
transformation~\cite{pl}. In Fig.~\ref{fig:spn}(a), we illustrate the
transformation in terms of $K$, which is indexed by the iteration step $n$.
The bare coupling $K$ will be hence identified with $K_0$. This block-spin
transformation replaces the upper spin block having two triangles and
intermediate bonds by a single bond, mapping $K_n$ to $K_{n+1}$. We consider
such a transformation at the outmost boundary layer so that an EBT with $L$
layers can be mapped to another EBT with $L-1$ layers. The block-spin
transformation can be performed by the majority rule at each triangle
without ambiguity. The
results are
\begin{eqnarray}
e^{g + K_{n+1}} &=& e^{5K + 3K_n} + 2e^{-K + 3K_n} + e^{-3K + 3K_n} +
2e^{3K + K_n}\nonumber\\
&& + 2e^{-3K + K_n} + 2e^{3K - K_n} + 2e^{K - K_n} +
2e^{-3K - K_n} + 2e^{K - 3K_n}\label{eq:e1a}\\
&\equiv& G_1(K, K_n),\nonumber\\
e^{g - K_{n+1}} &=& e^{3K + K_n} + 4e^{K + K_n} + 2e^{-K +
K_n}\nonumber\\
&& + e^{-5K + K_n} + 2e^{K - K_n} + 4e^{-K - K_n} + 2e^{-K - 3K_n}
\label{eq:e1b}\\
&\equiv& G_2(K, K_n),\nonumber
\end{eqnarray}
with a certain analytic function $g$, which appears as a consequence of
removing short length scales. Therefore, one obtains a recursion relation
\begin{equation}
K_{n+1} = \frac{1}{2} \ln\left[ \frac{G_1(K,K_n)}{G_2(K,K_n)} \right],
\label{eq:recur}
\end{equation}
from Eqs.~(\ref{eq:e1a}) and (\ref{eq:e1b}). We are interested in a fixed
point $K_n = K_{n+1} = K_\infty$. Note that this fixed point is always
stable, if it exists, since the slope at the crossing is less than $1$
[Fig.~\ref{fig:spn}(b)]. So it cannot be a point of phase separation, and we
should instead ask ourselves when the renormalized coupling strength
$K_\infty$ becomes infinite. It makes sense since it is equivalent to asking
when the renormalized occupation probability becomes one in the
bond-percolation problem~\cite{bnd}. The limiting $K_\infty$ diverges to
infinity when $K \rightarrow K_c \approx 0.465~810$ [Fig.~\ref{fig:spn}(b)].
However, this estimate corresponds to $T_c < 2.2$, which is not supported by
the MC data above. We have also tried working with a larger spin block, but
it hardly improves the estimate. In spite of such defects, which will be
commented on again later, this block-spin transformation may give a glimpse of
how to perform an RG analysis on this structure.
\begin{figure}
\includegraphics[width=0.33\textwidth]{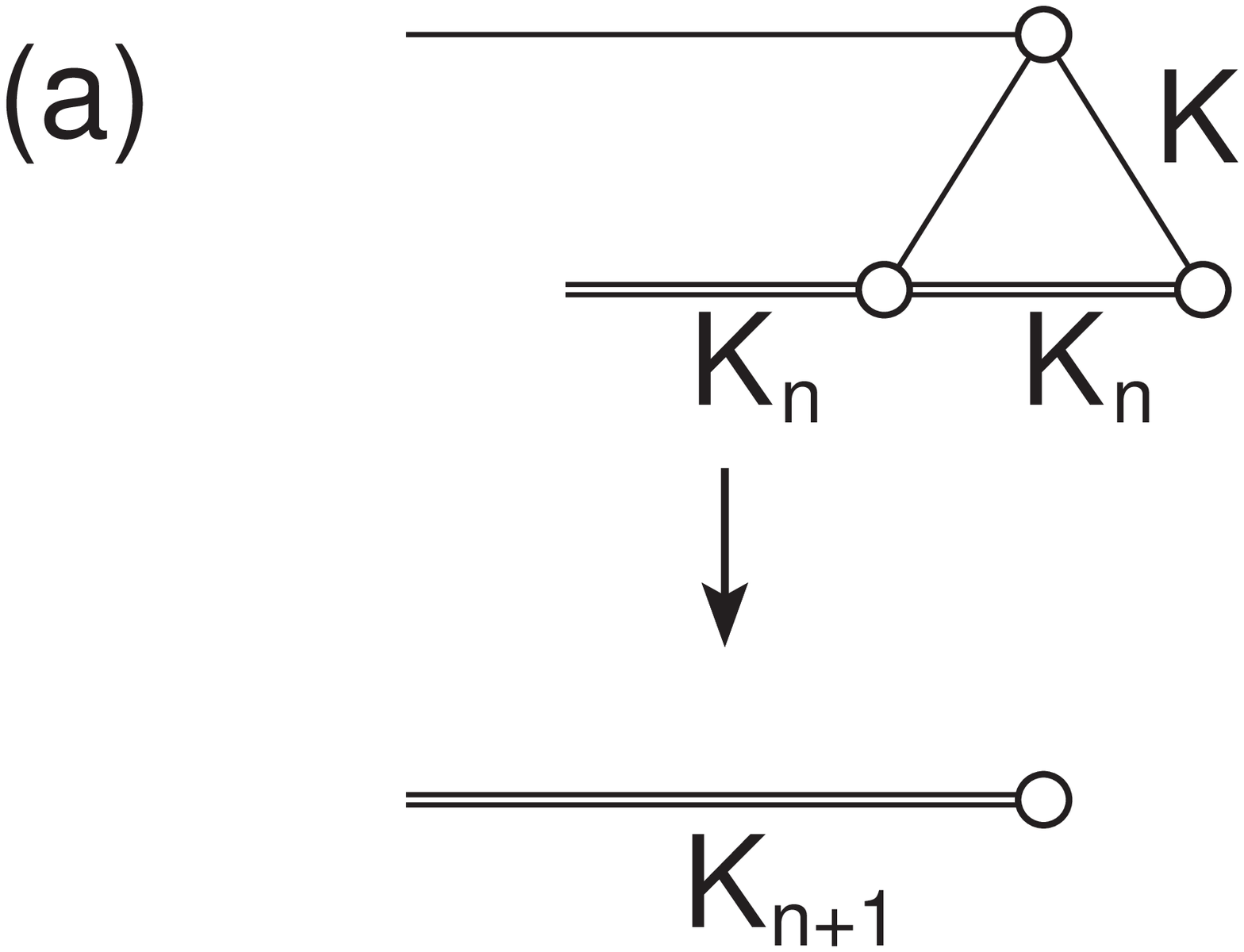}
\includegraphics[width=0.45\textwidth]{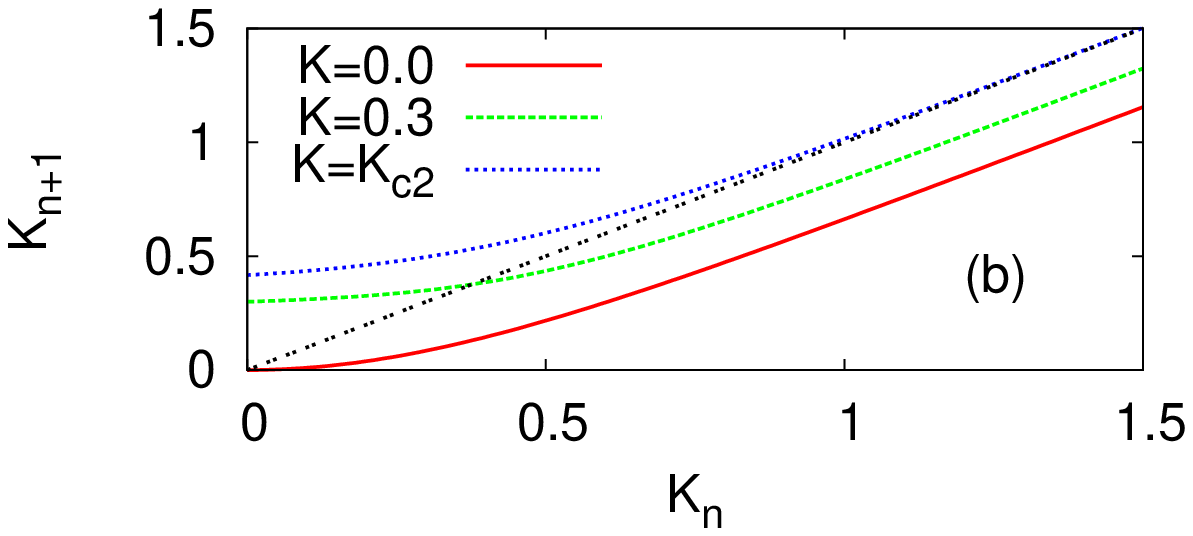}
\caption{(Color online) (a) Recursion scheme based on the transfer-matrix
method. The double lines represent coarse-grained effective couplings.
(b) Iteration by Eq.~(\ref{eq:two}) at various $K$'s.
The straight line has slope $1$, i.e., $K_{n+1} = K_n$.}
\label{fig:tm}
\end{figure}

In order to find a recursion scheme explaining the MC results better,
let us consider using transfer matrices. The idea is to describe exactly
a single layer first and then replace it by a 1D chain. It is
straightforward to write the transfer matrix for the 1D Ising chain with
coupling strength $K_{n+1}$ as
\begin{equation}
T = \left( \begin{array}{cc}
e^{K_{n+1}} & e^{-K_{n+1}}\\
e^{-K_{n+1}} & e^{K_{n+1}}
\end{array} \right),
\label{eq:mat}
\end{equation}
with two eigenvalues
$\lambda_1^{(n+1)} = e^{K_{n+1}} + e^{-K_{n+1}} = 2 \cosh K_{n+1}$ and
$\lambda_2^{(n+1)} = e^{K_{n+1}} - e^{-K_{n+1}} = 2 \sinh K_{n+1}$~\cite{pl}.
On the other hand, the upper spin block in Fig.~\ref{fig:tm}(a) has three
spins and therefore yields an $8 \times 8$ transfer matrix $T'$ in terms of
$K$ and $K_n$. The largest eigenvalue is obtained as
\[
\lambda_1^{(n)} = \frac{1}{2} e^{-3K-4K_n} \left[ 2e^{4K+2K_n} +
(e^{2K}+1)^2 e^{4K_n} + (e^{4K}+1) e^{2K+6K_n} + \sqrt{R_1(K,K_n)}
\right],
\]
with
\begin{eqnarray*}
R_1(K,K_n) &\equiv& \left[ (e^{2K_n}+2) (e^{2K} + e^{2K_n}) e^{2K+2K_n}
+ e^{6K+6K_n} + e^{4K_n} \right]^2\\
&& - 4(e^{4K}-1) (e^{4K_n}-1)^2 e^{4K+4K_n}.
\end{eqnarray*}
The second largest eigenvalue is also available as
\[
\lambda_2^{(n)} = \frac{1}{2} e^{-3K-4K_n} \left[ 2e^{4K+2K_n}
-(e^{2K}+1)^2 e^{4K_n} + (e^{4K}+1) e^{2K+6K_n} + \sqrt{R_2(K,K_n)}
\right],
\]
with
\begin{eqnarray*}
R_2(K,K_n) &\equiv& e^{6K+8K_n} \left\{ [-3\cosh K -\cosh 3K +
3\cosh(K-2K_n) + \cosh(3K+2K_n) \right.\\
&& \left. + 4\cosh K_n \cosh(K+K_n) \sinh 2K ]^2 - 32\sinh 2K
\sinh^2 2K_n \right\}.
\end{eqnarray*}
Since the matrix $T'$ has six more eigenvalues $\lambda_3^{(n)} >
\lambda_4^{(n)} > \cdots > \lambda_8^{(n)}$, we may need to consider eight
eigenvalues, or equivalently, eight energy levels from $E_1^{(n)} =
-\ln\lambda_1^{(n)}$ to $E_8^{(n)} = -\ln \lambda_8^{(n)}$ in total.
However, assuming that the first excitation from the ground state determines
the most dominant behavior, we approximate this with a two-level system
described by Eq.~(\ref{eq:mat}). According to this approximation, the first
energy gap $E_2^{(n)} - E_1^{(n)}$ between the two lowest levels should be
set equal to $E_2^{(n+1)} - E_1^{(n+1)} = -\ln
\left[\lambda_2^{(n+1)}/\lambda_1^{(n+1)} \right]$ in the simplified
two-level picture. In short, we keep the ratio between the two largest
eigenvalues at each iteration by
\begin{equation}
\frac{\lambda_2^{(n)}}{\lambda_1^{(n)}} =
\frac{\lambda_2^{(n+1)}}{\lambda_1^{(n+1)}},
\label{eq:two}
\end{equation}
which can be compared to Eq.~(\ref{eq:recur}) in our first attempt (see
Ref.~\cite{drell} where a similar idea is applied to the quantum Ising
chain). It can be also interpreted as adjusting correlation lengths since
the correlation length $\xi$ from a transfer-matrix calculation is given as
\begin{equation}
\xi^{(n)} = \frac{-1}{\ln \left[\lambda_2^{(n)}/ \lambda_1^{(n)}\right]}.
\label{eq:xi}
\end{equation}
Equation~(\ref{eq:two}) defines a recursion relation for getting $K_{n+1}$ out
of $K$ and $K_n$. We depict some cases of different $K$ values in
Fig.~\ref{fig:tm}(b). The limiting coupling strength $K_{\infty}$ becomes
infinite when $K \rightarrow K_c \approx 0.416~550~7$, which corresponds to
$T_c \approx 2.400~668$. Therefore, compared to the previous block-spin
transformation, this approach describes the MC data better.

From Eq.~(\ref{eq:xi}), we can see how the correlation length $\xi$ along
the boundary behaves
as $K$ approaches $K_c$. The behavior turns out to be
\begin{equation}
\xi = \frac{-1}{\ln \left[ \tanh K_{\infty}(K) \right]} \sim |K-K_c|^{-1/2}.
\label{eq:corr}
\end{equation}
It is related to correlation observed in a sufficiently inner part of the
system, but mediated by the boundary. The correlation-length exponent $\nu =
1/2$ in Eq.~(\ref{eq:corr}) indicates the mean-field result, which is
consistent with the prediction in Ref.~\cite{auriac} and the MC analysis in
Refs.~\cite{shima,sak}.
We again note that a number of boundary layers are mapped to a
1D chain that we are looking at, and that Eq.~(\ref{eq:corr}) is
obtained in this respect.
Although traveling along a single boundary layer is not the
shortest path between a pair of spins, the \emph{renormalized}
boundary does contain the shortest path so that the mean-field exponent is
found in this RG sense: it has been argued that the actual bare correlation
will be an exponentially decaying function due to the radius of
curvature~\cite{auriac}.
By introducing magnetic field $h_1$ at this renormalized boundary
part, we find
\[ \lambda_1 = e^{K_\infty} \cosh \beta h_1 + \sqrt{e^{2K_\infty} \sinh^2
\beta h_1 + e^{-2K_\infty}}, \]
which leads to the local susceptibility at $h_1 = 0$ as
\[ \chi_1 = \left. \frac{1}{\beta \lambda_1} \frac{\partial^2
\lambda_1}{\partial h_1^2} \right|_{h_1 = 0} = \beta e^{2K_\infty}. \]
From $K_\infty (K)$ shown in Eq.~(\ref{eq:corr}), it is straightforward
to find
\begin{equation}
\chi_1 \sim |K-K_c|^{-1/2}.
\label{eq:chi1}
\end{equation}
In the surface critical phenomena~\cite{henkel}, the correlation decays as
$G_\parallel(r) \sim r^{-d+2-\eta_\parallel}$ in the parallel direction to
the free surface, while it decays as $G_\perp(r) \sim r^{-d+2-\eta_\perp}$ in
the perpendicular direction. According to the RG theory, two-spin bulk
correlation $G(r) \sim r^{-d+2-\eta}$ is related to a scaling dimension
$x_h$ by $G(r) \sim r^{-2x_h}$, where $x_h = d - y_h$ with the RG eigenvalue
$y_h$. Suppose that the boundary scaling operator has
another scaling dimension $x_{h,s}$. Then $G_\parallel(r) \sim
r^{-2x_{h,s}}$ and $G_\perp(r) \sim r^{-x_h - x_{h,s}}$ depending on where
the spins lie, and it is therefore predicted that $\eta_\perp = (\eta
+ \eta_\parallel)/2$. The local susceptibility diverges as
$|K-K_c|^{-\gamma_1}$ with $\gamma_1 = \nu (2-\eta_\perp)$. At the upper
critical dimension, $\eta_\parallel = 2$ and $\eta = 0$, and the exponent
$\gamma_1$ is thus predicted to be $1/2$ in the mean-field theory.
Therefore, Eq.~(\ref{eq:chi1}) is fully consistent with the mean-field
description as well as Eq.~(\ref{eq:corr}). This mean-field result also
gives a clue to the poor performance of the block-spin transformation above:
the magnetization distribution gets more and more spread instead of having
sharp double peaks around $\pm 1$ as the dimension increases, so the concept
of a block spin becomes less adequate.

In summary, we have studied the Ising model on top of the EBT structure by
using the RG methods combined with the transfer-matrix calculation of its
single layer. It is an approximate calculation considering only the two
largest eigenmodes at each iteration, and it is hard to assess the error in
this approximation, as is usual in many real-space RG methods. We have
nevertheless reasonably predicted the order-disorder transition point, which
suggests that this approximate calculation can capture some essential
features of the system. Our main finding is that the critical behavior
predicted from this analysis is consistent with the mean-field theory of
free surfaces. Since many authors have expected that the bulk part far away
from the boundary will exhibit the bulk mean-field transition, our
RG analysis supports the mean-field picture all
the way up to the boundary part in understanding the Ising model on a
hyperbolic plane.

\acknowledgments
S.K.B. and P.M. acknowledge support from the Swedish Research Council
with Grant No. 621-2008-4449. 
B.J.K.  was supported by the Priority Research Centers Program
through the National Research Foundation of Korea (NRF)
funded by the Ministry of Education, Science, and Technology
(2010-0029700).


\end{document}